# AGFA-Net: Attention-Guided and Feature-Aggregated Network for Coronary Artery Segmentation using Computed Tomography Angiography

Xinyun Liu[1], Chen Zhao[2*]
1. Department of Computer Science, Michigan Technological University, Houghton, MI, 49931
2. Department of Computer Science, Kennesaw State University, Marietta, GA, 30062

Corresponding author:

Chen Zhao, PhD
Email address: czhao4@kennesaw.edu
Mailing address: 680 Arntson Dr, Atrium BLDG, Marietta, GA, 30060

**Abstract**:
Coronary artery disease (CAD) remains a prevalent cardiovascular condition, posing significant health risks worldwide. This pathology, characterized by plaque accumulation in coronary artery walls, leads to myocardial ischemia and various symptoms, including chest pain and shortness of breath. Accurate segmentation of coronary arteries from coronary computed tomography angiography (CCTA) images is crucial for diagnosis and treatment planning. Traditional segmentation methods face challenges in handling low-contrast images and complex anatomical structures. In this study, we propose an attention-guided, feature-aggregated 3D deep network (AGFA-Net) for coronary artery segmentation using CCTA images. AGFA-Net leverages attention mechanisms and feature refinement modules to capture salient features and enhance segmentation accuracy. Evaluation on a dataset comprising 1,000 CCTA scans demonstrates AGFA-Net's superior performance, achieving an average Dice coefficient similarity of 86.74% and a Hausdorff distance of 0.23 mm during 5-fold cross-validation. Ablation studies further validate the effectiveness of the proposed modules, highlighting their contributions to improved segmentation accuracy. Overall, AGFA-Net offers a robust and reliable solution for coronary artery segmentation, addressing challenges posed by varying vessel sizes, complex anatomies, and low image contrast.

## 1. Introduction
Coronary artery disease (CAD) remains the most prevalent cardiovascular condition and the leading cause of death globally [1]. CAD's pathology involves the accumulation of cholesterol-laden plaque within the coronary artery walls, causing varying degrees of stenosis. This stenosis reduces myocardial blood flow, and when significantly reduced, leads to myocardial ischemia, depriving the heart muscle of essential oxygen and nutrients. As the mismatch in blood supply worsens, symptoms such as chest pain (angina), shortness of breath, and other indicators of CAD manifest [2]. Additionally, unstable stenotic regions can acutely occlude the coronary artery, precipitating a heart attack [3]. Current treatments for significant coronary stenosis include percutaneous coronary intervention (PCI) [3] and coronary artery bypass grafting (CABG) [4], supplemented by rigorous medical therapies.
In clinical settings, coronary computed tomography angiography (CCTA) has emerged as a valuable non-invasive imaging technique for assessing coronary anatomy and guiding CAD diagnosis and treatment [5]. However, accurate automatic segmentation of coronary arteries from CCTA images poses several challenges:

1) low contrast between vessels and surrounding tissues; 2) motion artifacts due to cardiac and respiratory movements; 3) complex branching structures and varying vessel diameters. Manual assessment of coronary artery stenosis using CCTA is subjective and prone to inter-observer variability, leading to inconsistent evaluations of stenosis severity [6].

To address these challenges, various automatic segmentation techniques have been developed. Traditional image processing methods include three main categories: filter-based methods, line tracking-based methods, and model-based methods [7]. Filter-based methods typically utilize Gaussian filters [8] or Gabor filters [9] to enhance vessel edges and then apply adaptive thresholding to differentiate vessel pixels from the background. Line tracking methods start from seed points within the vessel and progressively identify vessel pixels based on intensity, contrast, and grayscale values [10,11]. Model-based methods employ active contours that balance external and internal forces to delineate vessel boundaries [12]. Despite their effectiveness, these traditional methods often struggle with low-contrast CCTA images.

In recent years, deep learning approaches, particularly convolutional neural networks (CNNs), have shown significant promise in automatic feature extraction and image segmentation. A CNN comprises multiple convolutional and pooling layers that hierarchically extract image features and utilize a classifier to distinguish between vessel and background pixels. For instance, Mu et al. designed a CNN to classify the central pixel of a local patch to segment coronary arteries in CCTA images [13]. Despite these advancements, further improvements in segmentation accuracy are necessary for reliable clinical application.

In our study, an attention-based method was developed to focus the convolutional neural network on the salient objects within the CCTA images, utilizing 1,000 annotated cases for coronary vascular tree extraction. We proposed a novel attention-guided, feature-aggregated 3D deep network (AGFA-Net) for coronary artery binary segmentation using these 1,000 publicly annotated CCTA datasets. Specifically, we introduced a feature refinement module for effective refinement of features during both the encoding and decoding stages, capturing latent semantic information effectively. Additionally, we proposed a scale-aware feature enhancement module aimed at dynamically adjusting the receptive field to extract more expressive features, thereby enhancing the network's feature representation capability. Moreover, we utilized a multi-scale feature aggregation module to learn more distinctive semantic representations, refining the vascular images. AGFA-Net, being an attention-guided and feature-aggregated network, achieved an average Dice coefficient similarity of 0.86 and a Hausdorff distance of 0.23 mm on the test cases during the 5-fold cross-validation.

## 2. Methodology

### 2.1. Network architecture

We propose a novel attention-guided, feature-aggregated 3D deep network, termed AGFA-Net, designed for accurate and reliable segmentation of coronary arteries. AGFA-Net is based on the classic U-shaped encoder-decoder architecture. The overall framework of AGFA-Net is illustrated in Figure 1. To enhance segmentation performance for challenging anatomical structures and shapes, such as irregular shapes, occlusions, scale variations, and blurred boundaries in lesions, we propose three practical modules: FRM, SAFA, and HFIM. Specifically, we propose the FRM (see Figure 2) to adaptively learn the fused spatial weights of feature maps at each scale. This module preserves detailed local semantic information while suppressing low-level background noise and irrelevant information. The proposed SAFA module captures long-range dependencies and effectively utilizes multi-channel space for feature representation and normalization. Additionally, it dynamically adjusts the receptive field, enabling the extraction of more expressive features. This enhances the network's capability to manage complex scenarios where the size and shape of coronary arteries vary significantly, and many arteries are intertwined. Finally, we present the HFIM module, which provides high-resolution, semantically strong feature vascular maps by extracting and fusing adjacent multi-level and multi-scale high-level features.

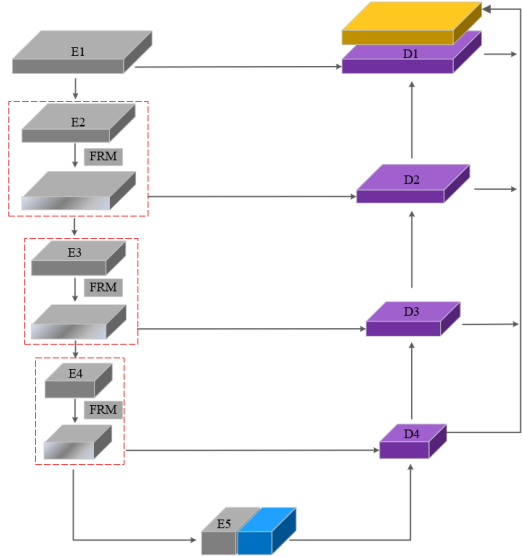

Figure 1. The architecture of AGFA-Net for coronary artery segmentation using CCTA.

## 2.2. Feature Refinement Module

To adaptively learn fused spatial weights of feature maps at each scale, retain detailed local semantic information, and suppress low-level irrelevant background noise, we have designed a Feature Refinement Module (FRM), as illustrated in Figure 2. This module employs both channel attention and spatial attention to enhance the expressive power of Convolutional Neural Networks (CNNs). The FRM captures feature $Y$ correlations by adaptively learning channel and spatial attention weights, thereby improving the performance of image feature extraction tasks. The FRM consists of two main components. Firstly, the channel attention mechanism gathers global statistical information for each channel using two pooling methods: global max pooling and global average pooling. This information is then processed by a shared two-layer neural network composed of a multi-layer perceptron (MLP) and a hidden layer. Unlike the SE method [14], which uses only the average pooling function, our approach also incorporates max pooling. This inclusion ensures that the maximum pooling feature $Y_{max}^c$, which encodes the degree of the most significant part, complements the average pooling feature $Y_{avg}^c$. Secondly, the spatial attention mechanism captures the maximum and average values at each spatial position using max pooling and average pooling. After applying these operations to the channels of each feature point, the resulting matrices are concatenated. The spatial position weights are then learned through a convolutional layer followed by a sigmoid function. These weights are finally applied to each spatial position on the feature map, resulting in features with enhanced spatial importance.

Given a feature map $Y$ as input, FRM sequentially infers a channel attention map $A_c$ and a spatial attention map $A_s$ as shown in Figure 2. To summarize, the attention process consists of:

$$Y' = A_c(Y) \otimes Y$$
$$Y'' = A_s(Y') \otimes Y' \quad (1)$$

where $\otimes$ indicating element-wise multiplication, and $Y''$ is the final refined output.

Feature Refine Module

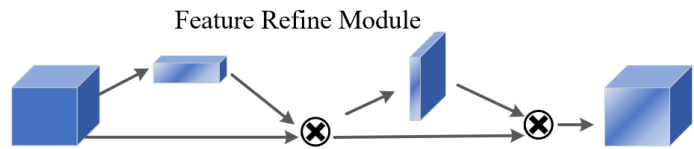

Figure 2. Architecture of FRM.

The channel attention is computed as:

$$A_c(Y) = f_{sig}\left(MLP(Y^c_{avg}) + MLP(Y^c_{max})\right) \quad (2)$$

where $f_{sig}$ denotes the sigmoid function, and $MLP$ indicates multi-layer perceptron with one hidden layer.
The computation of spatial attention is as follows:

$$A_s(Y) = f_{sig}\left(f^{7\times 7\times 7}([Y^s_{avg}; Y^s_{max}])\right) \quad (3)$$

where $f^{7\times 7\times 7}$ denotes a convolution operation with the filter size of $7\times 7\times 7$.

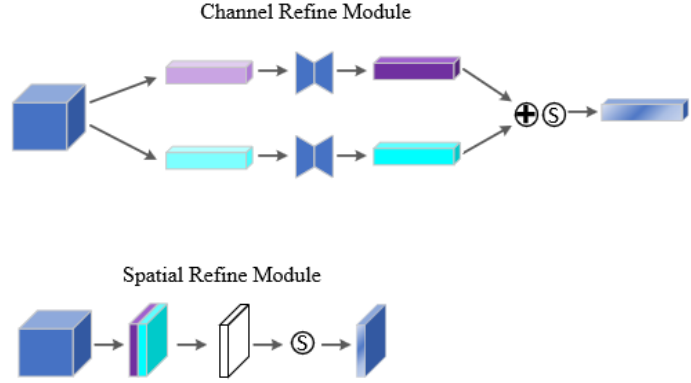

Figure 3. Architecture of scale-adaptive feature augmentation module.

### 3.3. Scale-adaptive Feature Augmentation Module

Due to the presence of small blood vessels around the coronary artery, coronary artery segmentation may produce fragmented blood vessels and incorrect segmentation. To address this issue, we propose a Scale-adaptive Feature Augmentation (ASFA) module, as illustrated in Figure 3, to effectively extract hidden multi-scale contextual information and enhance the network's feature representation capability. The ASFA module comprises two main components. The first component, Multi-resolution Feature Analysis, divides the generated feature maps into four parallel feature groups, indicated by $y_n$, $n \in \{1,2,3,4\}$. Each group $y_n$ retains the same spatial dimensions as the input feature but has one-fourth the number of channels. We apply dilated convolutions with varying dilation rates to these four groups, expanding the network's receptive field without losing spatial resolution. This approach enables the network to better capture contextual information in the image and produce richer feature maps with diverse receptive fields, allowing for comprehensive perception of features at different scales in CAS. The feature refinement process of the parallel features is as follows:

$$y'_n = f_{sig}(f_c^{d_n}(y_n)) \otimes f_c^{d_n}(y_n) \quad (4)$$

where $f_c^{d_n}$ denotes the dilated convolution layer with dilated rate of $d_n$ ($d_1 = 1$, $d_2 = 2$, $d_3 = 3$, and $d_4 = 4$ in this paper); $f_{sig}$ indicates the sigmoid function. Next, the features $Y$ from these different channels are concatenated and passed to the second component, **Dynamic Feature Identification**, which further accesses the importance of each feature.

$$Y = f_{con}(y'_1, y'_2, y'_3, y'_4) \quad (4)$$

where $f_{con}$ represents the concatenation operation.
Based on the attention mechanism from the Transformer model, we can effectively extract and focus on relevant features. Initially, the hierarchical features $Y$ are processed through three convolutional layers, each followed by batch normalization (BN) and a ReLU activation function. The convolutional layers have kernel sizes $3\times 1\times 1$, $1\times 3\times 1$, and $1\times 1\times 3$ respectively, and generate three feature maps $Q \in \mathbb{R}^{C\times D\times H\times W}$, $K \in \mathbb{R}^{C\times D\times H\times W}$, $V \in$

$\mathbb{R}^{C \times D \times H \times W}$ (where $C$, D, $H$ and $W$ denote the channel, depth, height and width of the input features $Y$, respectively). Subsequently, we conduct matrix multiplication of $Q^T$ and $K$ to encode the feature relationships. Following that, we obtain the result by performing matrix multiplication between it and $V$, which is a context-aware representation of the features $Y$, as the output vector has been constructed by focusing on the most relevant parts of the entire input. Moreover, a residual connection is established to integrate the multi-scale feature maps, and the expression of the SAFA module output is formulated as follows:

$$Y_{SAFA} = Y + f_{softmax}(Q^T \otimes K) \otimes V \tag{5}$$

where $\otimes$ means matrix multiplication. Through the incorporation of the SAFA module, appended subsequent to the final layer $E_5$ of the encoder, the network's capacity for feature representation is significantly augmented.

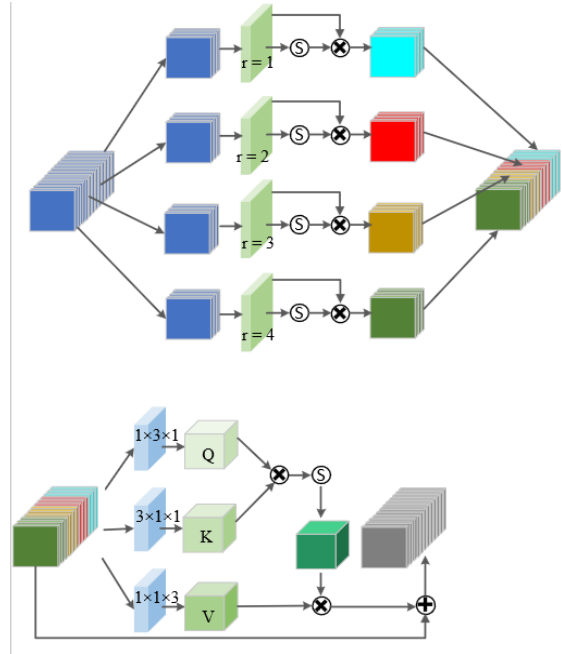

Figure 4. Architecture of SAFA module.

**2.4 Hierarchical Feature Integration Module**

To meet the segmentation requirements for identifying various sizes of CA from complex backgrounds, it is essential to integrate multiple feature types. Liu et al. demonstrate that aggregating multi-scale features effectively enhances image resolution [15]. Similarly, Li et al. show that this aggregation structure can improve feature representation capabilities [16]. Unlike previous works that directly predict refined features through multi-branch prediction, we propose a HFIM module, as illustrated in Figure 5, to gradually aggregates adjacent scale features in the decoding path to obtain more detailed semantic information.

Specifically, the features $Y_n$, $n \in \{1,2,3,4\}$ between two adjacent scales are concatenated in a right-to-left direction. A soft attention map is then used to guide the effective expression of CA features across different scales. This Feature Fusion Upsample approach allows for the extraction of multi-scale deep semantic information, enhancing both spatial and locational information.

$$\begin{aligned} Y_3^I &= f_c \left( f_{sig} \left( \mathbb{C} \left( Y_3, f_u(Y_4) \right) \right) \right) \otimes Y_3 \\ Y_2^I &= f_c \left( f_{sig} \left( \mathbb{C} \left( Y_2, f_u(Y_3) \right) \right) \right) \otimes Y_2 \\ Y_1^I &= f_c \left( f_{sig} \left( \mathbb{C} \left( Y_1, f_u(Y_2) \right) \right) \right) \otimes Y_1 \end{aligned} \tag{6}$$

where $f_u$ refers the upsampling operation; $\mathbb{C}$ represents the concatenation operation; $\otimes$ represents the matrix multiplication; $f_{sig}$ indicates the sigmoid activation function used to obtain the attention map at the current scale; and $f_c$ applies a convolution with the $3 \times 3 \times 3$ kernel to generate features with rich, detailed information.

The HFIM module is used to interact with different scale features to enhance the semantic information of the pixels and incrementally guide feature aggregation between adjacent scales.

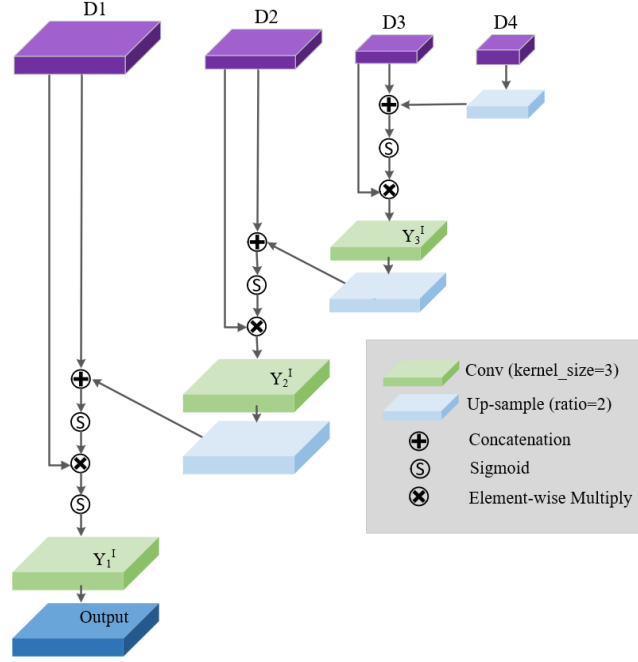

Figure 5. Overview of the HFIM.

## 2.5. Loss function

Coronary artery segmentation in CCTA images can be regarded as a pixel-level binary classification task: coronary artery or background. When the segmentation process focuses on rare observations, it often results in a significant class imbalance among the candidate labels, leading to suboptimal performance. To mitigate this issue, various strategies have been proposed, including the weighted cross-entropy function, the sensitivity function, and the Dice loss function [17]. Consequently, we selected to utilize the weighted cross-entropy function and the Dice loss function as our loss functions to adjust the learning bias between vessels and the background during training. Therefore, we define the 3D optimization loss function for training the proposed AGFA-Net as follows:

$$\mathcal{L} = \lambda L_{WCE} + (1 - \lambda) L_{Dice} \quad (7)$$

$$L_{Dice} = 1 - \frac{2 \sum_{n=1}^{N} g_n \cdot p_n + \epsilon}{\sum_{n=1}^{N} g_n + \sum_{n=1}^{N} p_n + \epsilon} \quad (8)$$

$$L_{WCE} = -\frac{1}{N} \sum_{n=1}^{N} (\omega g_n \log p_n + (1 - g_n) \log(1 - p_n)) \quad (9)$$

$$\omega = \frac{N - \sum_{n=1}^{N} p_n}{\sum_{n=1}^{N} p_n} \quad (10)$$

where $N$ represents the number of voxels, and $p_n \in [0,1]$ and $g_n \in [0,1]$ indicate the predicted probability and ground truth value of the $n^{th}$ voxel as the vessel structure, respectively. Here the weight balance parameter $\lambda$ between the WCE loss and the Dice loss is empirically set to 0.6. The parameter $\epsilon$ serves as a Laplace smoothing factor employed to mitigate numerical instability issues and expedite the convergence of the training process (set to $\epsilon = 1.0$ in this study); $\omega$ represents the weight attributed to the vessel structure, which is determined based on the class estimation probabilities $p_n$ of all voxels.

## 3. Experiments and Results
### 3.1 Dataset

We evaluate our proposed method using an in-house dataset comprising CCTA data, retrospectively collected from 1,000 patient cases, for medical volumetric segmentation. The data were acquired using multiple CT scanners, specifically the Revolution CT by GE Healthcare and the SOMATOM Definition Flash by Siemens Healthcare [18]. All images were obtained at the cardiovascular medicine department of a top-grade (Grade III Level A) hospital in the country, with scanning conducted by certified radiology technicians. Ethical approval for all cases was granted by the institutional review board. Three experienced radiologists annotated the data using 3D Slicer, a free, open-source software for medical image processing.

Detailed characteristics of the dataset are described below. This dataset comprises 3D CCTA scans with pixel spacing ranging from 0.28 to 0.41 mm. The slice thickness varies between 0.5 and 1.0 mm, and the number of slices per scan ranges from 199 to 275. The patients' ages in the dataset span from 46 to 78 years.

### 3.2. Experimental settings and evaluation criteria

*Experimental Settings.* The proposed network is implemented using the PyTorch library and is trained on four Tesla V100-PCIE graphic cards, each with 32 GB of memory. We employ adaptive moment estimation (Adam) as the optimizer. The initial learning rate is set to 0.003, with a weight decay of $1 \times 10^{-6}$, and a CosineAnnealingWarmRestarts learning rate policy is applied. All networks are trained for 500 epochs, and we set the batch size as 16. To achieve peak performance for AGFA-Net, data augmentation is performed. The data augmentation strategy includes three random operations: rotation within -20 to 20 degrees, horizontal flipping, and cropping. Moreover, we randomly cropped subvolumes of size $128 \times 160 \times 160$ from the input data for model training.

To ensure fair and reliable performance evaluation, five-fold cross-validation is conducted on the CCTA dataset, and the average performance across all evaluation criteria is reported. To accelerate the training process and enhance model performance, we normalize the volume data, including both training and test sets, to conform to a specific range or distribution. The CCTA dataset consists of 800 volumes for training and 200 volumes for testing. During the training phase, 120 volumes (15% of the training data) are randomly selected to serve as the validation set.

*Evaluation Criteria.* To evaluate the performance of different models, we employ three widely used criteria: Dice, Recall, and Precision. These metrics range from 0 to 1, with higher values indicating better segmentation accuracy. The specifics of each criterion are as follows:

$$Dice = \frac{2 \cdot TP}{(TP + FN) + (TP + FP)} \qquad (11)$$

$$Recall = \frac{TP}{TP + FN} \qquad (12)$$

$$Precision = \frac{TP}{TP + FP} \qquad (13)$$

where TP, TN, FP, FN are True Positive, True Negative, False Positive and False Negative, respectively. Dice was considered the most important criterion for segmentation comparison, and participants were ranked based on this metric [19]. Consequently, when quantitatively evaluating the performance of a network, we prioritize Dice, followed by Recall and Precision.

### 3.3. Comparisons with the state-of-the-art methods

In this subsection, we compare our proposed method with two state-of-the-art 3D segmentation networks: U-Net3D and V-Net. We evaluate the performance of each method on our CCTA dataset using evaluation criteria, and we discuss the strengths and weaknesses of each approach. U-Net3D is an extension of the 2D U-Net architecture designed for volumetric segmentation tasks [20]. It employs a symmetrical encoder-decoder structure with skip connections that enable the network to capture both fine and coarse details from the input volume. V-Net is another

3D segmentation network that uses a similar encoder-decoder architecture but introduces residual connections [21], which helps to alleviate the vanishing gradient problem and allows for more efficient training.

### 3.3.1. Quantitative results

Table 1 quantitatively shows the performance of AGFA-Net and the two comparison methods on the CCTA dataset. All evaluation criteria in Table 1 are obtained by averaging the five-fold cross validation. Experimental results demonstrate that our AGFA-Net outperforms other networks in coronary artery segmentation.

Specifically, the Dice coefficient, which measures the overlap between the predicted segmentation and the ground truth, shows that AGFA-Net outperforms the other networks with a score of 86.74%, which is 5.68% higher than U-Net3D and 3.31% higher than V-Net. This indicates that AGFA-Net provides the most accurate segmentation results among the three networks. Recall, which indicates the ability of the model to identify all relevant instances, is highest for AGFA-Net at 98.79%. AGFA-Net's recall is 1.65% higher than V-Net and 2.12% higher than U-Net3D. The higher recall of AGFA-Net suggests that it is more effective at detecting true positive cases. Precision, which measures the accuracy of the positive predictions, is also highest for AGFA-Net at 90.53%. AGFA-Net's precision is 5.26% higher than V-Net and 7.05% higher than U-Net3D. This demonstrates that AGFA-Net not only detects more relevant instances but also makes fewer false positive predictions.

**Table 1**. Performance of coronary artery segmentation between the proposed AGFA-Net and existing methods.

| Network | Dice | Recall | Precision |
|---|---|---|---|
| U-Net3D | 81.06% | 96.67% | 83.48% |
| V-Net | 83.43% | 97.14% | 85.27% |
| AGFA-Net | 86.74% | 98.79% | 90.53% |

### 3.3.2. Qualitative results

The qualitative experimental results of our method, along with those of other competitors on typical cases, are shown in Figs. 2. After performing the 3D morphological closing operation and the process of preserving the maximum connected region, the vessel surface becomes smoother, and some noise blocks are eliminated. We employ k3d's toolbox to visualize coronary artery segmentation results and compare qualitative outcomes. As shown, our method yields more accurate and visually coherent segmentations, particularly in regions with complex anatomical structures. For coronary artery branches, competitors often produce fragments that do not belong to the coronary artery and frequently lose coronary artery details.

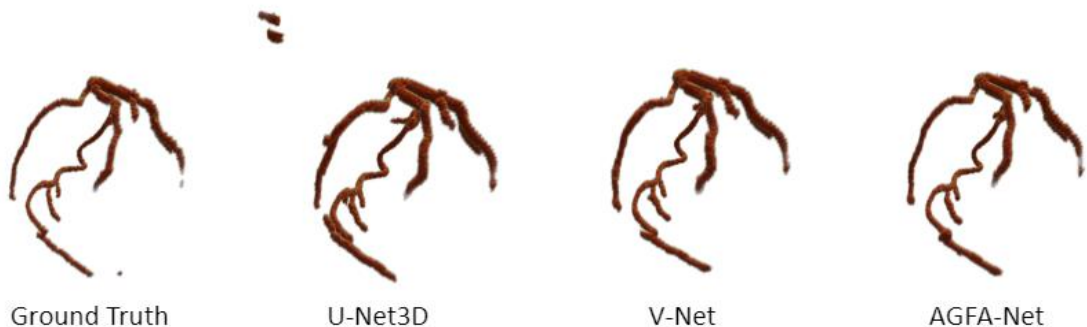

Figure 6. Visual comparison of the proposed AGFA-Net and existing STOA networks for coronary artery segmentation.

Through both quantitative and qualitative analysis, it is evident that the proposed AGFA-Net delivers a significant and consistent performance improvement. This is attributed to the synergistic integration of the FRM, SAFA, and

HFIM modules. These findings demonstrate AGFA-Net's effectiveness in addressing the large-scale variations and complex semantics of CA in this challenging task.

### 3.4. Ablation studies

In this section, we conduct an ablation study to evaluate the effectiveness of each module in the AGFA-Net: FRM, SAFA, and HFIM. We assess the contribution of these modules to the overall performance of our network on the task of coronary artery segmentation.

**FRM:** The Feature Refinement Module (FRM) is designed to enhance the feature representation by refining the extracted features. To determine the impact of the FRM, we compare the performance of networks with and without the FRM. As illustrated in table 2, Net 1, Net 7, Net 8, and AGFA-Net include the FRM. The results show that the inclusion of the FRM leads to a notable improvement in performance metrics. For instance, Net 1 achieves a Dice score of 83.31%, compared to the Baseline's 79.91%. The enhancement in Dice, Recall, and Precision values across these networks demonstrates the significant role of the FRM in improving segmentation accuracy. To further validate the effectiveness of the FRM module, we compared the output of Baseline and Net1 using 3D visualization. As illustrated in Fig.7, Net1 captures a more complete topology and better suppresses background noise compared to the Baseline.

**SAFA:** The Scale-Adaptive Feature Augmentation (SAFA) module is crucial for aggregating features with a self-attention mechanism to capture long-range dependencies. The effectiveness of SAFA is highlighted by comparing the networks that incorporate this module (Net 2, Net 3, Net 4, Net 5, Net 7, Net 9, and AGFA-Net) with those that do not. For example, Net 4 shows an increase in Dice score to 84.27%, which has the best dilated convolution rate compared with Net 2, Net3 and Net 5, and it increases by 4.36% compared with the baseline. The improvement underlines the importance of self-attention in enhancing the segmentation capability by effectively aggregating relevant features. Fig. 7 demonstrates that Net 4, in comparison to the Baseline, can segment more under-segmented vessels, particularly those with smaller scales. This highlights that the dynamic selection of multi-scale contextual information significantly enhances vessel segmentation.

**HFIM:** The Hierarchical Feature Integration Module (HFIM) is designed to gradually aggregates adjacent scale features in the decoding path to obtain more detailed semantic information and better capture multi-scale information. To analyze its contribution, we compare the networks that utilize the HFIM (Net 6, Net 8, Net 9, and AGFA-Net) against those that do not. Networks incorporating HFIM consistently perform better, as seen in Net 6 with a Dice score of 84.14% and Net 8 with 85.09%. AGFA-Net, which includes all three modules, achieves the highest Dice score of 86.74%, demonstrating the synergistic effect of combining these modules. As illustrated in Fig. 7, a comparison between Net 7 (Baseline + FRM + SAFA) and AGFA-Net (Baseline + FRM + SAFA + HFIM) reveals that AGFA-Net achieves superior segmentation results. This improvement is attributed to the HFIM module's capability to extract more semantic and hierarchical information.

**Overall Performance:** The integration of FRM, SAFA, and HFIM in AGFA-Net results in significant and consistent performance benefits. AGFA-Net achieves the highest metrics with a Dice score of 86.74%, Recall of 98.79%, and Precision of 90.53%. This confirms the effectiveness of our proposed modules in addressing the challenges of large-scale variations and complex semantics in coronary artery segmentation tasks. The results validate the advantage of seamlessly integrating FRM, SAFA, and HFIM, thereby proving the robustness and superiority of AGFA-Net.

To further evaluate the effectiveness of the proposed modules, we initially integrate FRM+SAFA, FRM+HFIM, and SAFA+HFIM into the baseline model, resulting in Net 7, Net 8, and Net 9, respectively. Subsequently, we combine all three modules, FRM, SAFA, and HFIM, into the Baseline, creating AGFA-Net. As depicted in Table 2, the performance of both Net 7 and Net 8 significantly surpasses that of Net 1 (Baseline + FRM). Illustrated in Fig. 7, the vascular segmentation produced by Net 7 is noticeably smoother and more complete compared to Net 1. These findings demonstrate the complementary nature of the FRM and SAFA (or HFIM) modules, yielding more consistent and complete vascular segmentation outcomes. According to Table 2, AGFA-Net shows marked improvement over the Baseline, Net 1, Net 4, and Net 6. Specifically, the Dice for AGFA-Net increased by 6.83% compared to the Baseline, a substantial improvement over the gains observed in Net 1 ( ↑ 3.40%), Net 4 ( ↑ 4.36%), and Net 6 ( ↑ 4.23%). This highlights the synergistic effects achieved by integrating FRM, SAFA, and HFIM.

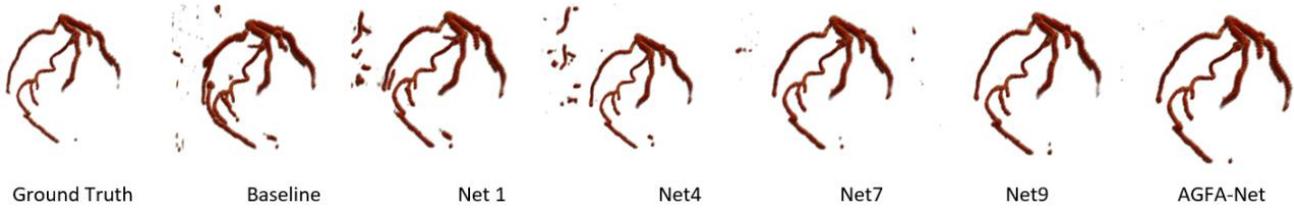

Figure 7. Visualization of the coronary artery segmentation results using baseline models and ablation studies.

Table 2. Ablation study of the proposed AGFA-Net for coronary artery segmentation.

| Network | FRM | SAFA | | | | HFIM | self-attention | Dice | Recall | Precision |
| --- | --- | --- | --- | --- | --- | --- | --- | --- | --- | --- |
| | | D1 | D2 | D3 | D4 | | | | | |
| Baseline | | | | | | | | 79.91% | 96.58% | 83.13% |
| Net 1 | √ | | | | | | √ | 83.31% | 96.85% | 85.22% |
| Net 2 | | √ | | | | | √ | 84.03% | 97.02% | 85.74% |
| Net 3 | | | √ | | | | √ | 84.22% | 97.32% | 86.05% |
| Net 4 | | | | √ | | | √ | 84.27% | 97.48% | 87.79% |
| Net 5 | | | | | √ | | √ | 84.19% | 97.74% | 86.63% |
| Net 6 | | | | | | √ | √ | 84.14% | 97.68% | 86.44% |
| Net 7 | √ | | | √ | | | √ | 85.23% | 98.54% | 89.27% |
| Net 8 | √ | | | | | √ | √ | 85.09% | 98.28% | 88.89% |
| Net 9 | | | | √ | | √ | √ | 85.41% | 98.45% | 89.36% |
| AGFA-Net | √ | | | √ | | √ | √ | 86.74% | 98.79% | 90.53% |

## 4. Discussion

Coronary artery segmentation from CCTA images plays a pivotal role in CAD diagnosis and treatment planning. Our study proposes AGFA-Net, a novel attention-guided, feature-aggregated 3D deep network, for automatic segmentation of coronary arteries. AGFA-Net leverages attention mechanisms and feature refinement modules to enhance segmentation accuracy, addressing challenges posed by low image contrast and complex anatomical structures.

The performance evaluation of AGFA-Net demonstrates its superiority over traditional segmentation methods and state-of-the-art deep learning networks. With an average Dice coefficient similarity of 86.74% and a Hausdorff distance of 0.23 mm, AGFA-Net achieves robust segmentation results, outperforming existing approaches. The substantial improvement in segmentation accuracy can be attributed to the synergistic integration of the proposed modules: Feature Refinement Module, Scale-Adaptive Feature Augmentation module, and Hierarchical Feature Integration Module. These modules enhance feature representation, capture multi-scale contextual information, and facilitate effective feature aggregation, thereby improving the network's segmentation capability.

Ablation studies further validate the effectiveness of each module in AGFA-Net. The FRM significantly enhances segmentation accuracy by refining extracted features and suppressing background noise. The SAFA module effectively aggregates multi-scale contextual information, addressing challenges posed by small blood vessels and complex anatomical structures. The HFIM further improves segmentation accuracy by gradually aggregating adjacent scale features, enabling the network to capture detailed semantic information.

The robustness and reliability of AGFA-Net make it a promising tool for CAD diagnosis and treatment planning. By accurately segmenting coronary arteries from CCTA images, AGFA-Net can assist clinicians in assessing disease severity, guiding treatment decisions, and monitoring disease progression. Moreover, AGFA-Net's ability to handle variations in vessel sizes, complex anatomies, and low image contrast enhances its applicability in real-world clinical settings.

While our study demonstrates significant advancements in coronary artery segmentation, several avenues for future research exist. Further exploration of attention mechanisms and feature refinement techniques could enhance AGFA-Net's segmentation performance. Additionally, investigating the generalizability of AGFA-Net across diverse patient populations and imaging modalities could broaden its clinical utility. Overall, AGFA-Net represents a promising step towards accurate and reliable coronary artery segmentation, contributing to improved CAD diagnosis and patient care.

## 5. Conclusion

In summary, our study introduces AGFA-Net, a novel attention-guided, feature-aggregated 3D deep network, tailored for automatic segmentation of coronary arteries from CCTA images. With CAD remaining a foremost cardiovascular concern globally, the accurate delineation of coronary anatomy is pivotal for precise diagnosis and treatment planning. AGFA-Net addresses this imperative by harnessing attention mechanisms and advanced feature refinement modules to significantly enhance segmentation accuracy. Through extensive experimentation and comparative analysis, AGFA-Net demonstrates superior performance over traditional methods and contemporary deep learning architectures, achieving robust segmentation results with an average Dice coefficient similarity of 86.74% and a Hausdorff distance of 0.23 mm. The integration of key modules, including the Feature Refinement Module FRM, Scale-Adaptive Feature Augmentation SAFA module, and Hierarchical Feature Integration Module HFIM, amplifies AGFA-Net's segmentation prowess, promising improved CAD diagnosis and patient care in clinical settings.

**Reference**

[1] Okrainec K, Banerjee DK, Eisenberg MJ. Coronary artery disease in the developing world. American heart journal. Elsevier; 2004;148(1):7–15.

[2] Costa PT. Influence of the normal personality dimension of neuroticism on chest pain symptoms and coronary artery disease. The American Journal of Cardiology. 1987 Dec;60(18):J20–J26.

[3] Hamada S, Kashiwazaki D, Yamamoto S, Akioka N, Kuwayama N, Kuroda S. Impact of Plaque Composition on Risk of Coronary Artery Diseases in Patients with Carotid Artery Stenosis. Journal of Stroke and Cerebrovascular Diseases. 2018 Dec;27(12):3599–3604.

[4] Acharjee S, Teo KK, Jacobs AK, Hartigan PM, Barn K, Gosselin G, Tanguay J-F, Maron DJ, Kostuk WJ, Chaitman BR, Mancini GBJ, Spertus JA, Dada MR, Bates ER, Booth DC, Weintraub WS, O'Rourke RA, Boden WE. Optimal medical therapy with or without percutaneous coronary intervention in women with stable coronary disease: A pre-specified subset analysis of the Clinical Outcomes Utilizing Revascularization and Aggressive druG Evaluation (COURAGE) trial. American Heart Journal. 2016 Mar;173:108–117.

[5] Song A, Xu L, Wang L, Wang B, Yang X, Xu B, Yang B, Greenwald SE. Automatic Coronary Artery Segmentation of CCTA Images With an Efficient Feature-Fusion-and-Rectification 3D-UNet. IEEE J Biomed Health Inform. 2022 Aug;26(8):4044–4055.

[6] Williams MC, Golay SK, Hunter A, Weir-McCall JR, Mlynska L, Dweck MR, Uren NG, Reid JH, Lewis SC, Berry C, Van Beek EJR, Roditi G, Newby DE, Mirsadraee S. Observer variability in the assessment of CT coronary angiography and coronary artery calcium score: substudy of the Scottish COmputed Tomography of the HEART (SCOT-HEART) trial. Open Heart. 2015 May;2(1):e000234.

[7] Zhao C, Vij A, Malhotra S, Tang J, Tang H, Pienta D, Xu Z, Zhou W. Automatic extraction and stenosis evaluation of coronary arteries in invasive coronary angiograms. Computers in Biology and Medicine. Elsevier; 2021;136:104667.

[8] Li Y, Zhou S, Wu J, Ma X, Peng K. A novel method of vessel segmentation for X-ray coronary angiography images. 2012 Fourth International Conference on Computational and Information Sciences. IEEE; 2012. p. 468–471.

[9] Felfelian B, Fazlali HR, Karimi N, Soroushmehr SMR, Samavi S, Nallamothu B, Najarian K. Vessel segmentation in low contrast X-ray angiogram images. 2016 IEEE International Conference on Image Processing (ICIP). IEEE; 2016. p. 375–379.

[10] Vlachos M, Dermatas E. Multi-scale retinal vessel segmentation using line tracking. Computerized Medical Imaging and Graphics. Elsevier; 2010;34(3):213–227.

[11] Guerrero J, Salcudean SE, McEwen JA, Masri BA, Nicolaou S. Real-time vessel segmentation and tracking for ultrasound imaging applications. IEEE transactions on medical imaging. IEEE; 2007;26(8):1079–1090.

[12] Taghizadeh Dehkordi M, Doost Hoseini AM, Sadri S, Soltanianzadeh H. Local feature fitting active contour for segmenting vessels in angiograms. IET Computer Vision. Wiley Online Library; 2014;8(3):161–170.

[13] Mu N, Lyu Z, Rezaeitaleshmahalleh M, Tang J, Jiang J. An attention residual u-net with differential preprocessing and geometric postprocessing: Learning how to segment vasculature including intracranial aneurysms. Medical Image Analysis. 2023 Feb;84:102697.

[14] Hu J, Shen L, Albanie S, Sun G, Wu E. Squeeze-and-Excitation Networks [Internet]. arXiv; 2019 [cited 2024 Mar 30]. Available from: http://arxiv.org/abs/1709.01507

[15] Liu J, Zhang W, Tang Y, Tang J, Wu G. Residual feature aggregation network for image super-resolution. Proceedings of the IEEE/CVF conference on computer vision and pattern recognition [Internet]. 2020 [cited 2024 Jun 11]. p. 2359–2368. Available from: http://openaccess.thecvf.com/content_CVPR_2020/html/Liu_Residual_Feature_Aggregation_Network_for_Image_

Super-Resolution_CVPR_2020_paper.html

[16] Li H, Xiong P, Fan H, Sun J. Dfanet: Deep feature aggregation for real-time semantic segmentation. Proceedings of the IEEE/CVF conference on computer vision and pattern recognition [Internet]. 2019 [cited 2024 Jun 11]. p. 9522–9531. Available from: http://openaccess.thecvf.com/content_CVPR_2019/html/Li_DFANet_Deep_Feature_Aggregation_for_Real-Time_Semantic_Segmentation_CVPR_2019_paper.html

[17] Sudre CH, Li W, Vercauteren T, Ourselin S, Jorge Cardoso M. Generalised dice overlap as a deep learning loss function for highly unbalanced segmentations. Deep learning in medical image analysis and multimodal learning for clinical decision support. Springer; 2017. p. 240–248.

[18] Dong C, Xu S, Dai D, Zhang Y, Zhang C, Li Z. A novel multi-attention, multi-scale 3D deep network for coronary artery segmentation. Medical Image Analysis. 2023 Apr;85:102745.

[19] Codella N, Rotemberg V, Tschandl P, Celebi ME, Dusza S, Gutman D, Helba B, Kalloo A, Liopyris K, Marchetti M, Kittler H, Halpern A. Skin Lesion Analysis Toward Melanoma Detection 2018: A Challenge Hosted by the International Skin Imaging Collaboration (ISIC) [Internet]. arXiv; 2019 [cited 2024 Jun 11]. Available from: http://arxiv.org/abs/1902.03368

[20] Zeng G, Yang X, Li J, Yu L, Heng P-A, Zheng G. 3D U-net with Multi-level Deep Supervision: Fully Automatic Segmentation of Proximal Femur in 3D MR Images. Cham: Springer International Publishing; 2017. p. 274–282.

[21] Milletari F, Navab N, Ahmadi S-A. V-net: Fully convolutional neural networks for volumetric medical image segmentation. 2016 fourth international conference on 3D vision (3DV). IEEE; 2016. p. 565–571.